\documentstyle[prl,aps,epsf,multicol]{revtex}
\begin{document}
\draft

\title{Conductance of the Single Electron Transistor for Arbitrary
       Tunneling Strength}
\author{Georg G\"oppert$^1$, Bruno H\"upper$^2$, and Hermann Grabert$^1$}
\address{$^1$ Fakult\"at f\"ur Physik, Albert--Ludwigs--Universit{\"a}t, \\
Hermann--Herder--Stra{\ss}e~3, D--79104 Freiburg, Germany \\
$^2$ Chemical Physics Department, Weizmann Institute of Science,
76100 Rehovot, Israel}

\date{\today}
\maketitle
\widetext

\begin{abstract}
We study the temperature and gate voltage dependence of the
conductance of the single electron transistor  
focusing on highly conducting devices.
Electron tunneling is treated nonperturbatively by means of 
path integral Monte Carlo techniques and the conductance is 
determined from the Kubo formula. 
A regularized singular value decomposition scheme is employed to
calculate the conductance from imaginary time simulation data.
Our findings are shown to bridge between available analytical results 
in the semiclassical and perturbative limits and are found to explain 
recent experimental results in a regime not accessible by earlier
methods. 
\end{abstract}


\raggedcolumns
\begin{multicols}{2}
\narrowtext

\vspace*{-0.6cm}

The single electron transistor (SET) has become a primary device for
applications of the Coulomb blockade effect \cite{Nato92}.
The modulation of the current in dependence on the gate voltage 
has been used to build highly sensitive electrometers and thermometers
and can serve as switch or amplifier. Since the relative current 
modulation is most pronounced at small transport voltages 
where also heating effects become negligible, the quantity of 
chief interest is the linear conductance. 
There are two relevant dimensionless parameters \cite{Nato92}: 
the dimensionless parallel conductance 
$g$ of the two tunnel junctions measured in units of the conductance
quantum $G_K=e^2/h$, and the dimensionless inverse temperature
$\beta E_C$ relating $\beta=1/k_B T$ with the charging energy $E_C$. 
While recent experiments 
\cite{KastnerRMP92,JoyezSETPRL97,KuzminSETPRB99}
have explored a large regime
of tunneling conductances and temperatures, available
theoretical work is restricted to 
limiting cases where one of the parameters $g$ or 
$\beta E_C$ is small. Approaches for the weakly
conducting regime 
$g {\tiny{\lower 2pt \hbox{$<$}\atop\raise 5pt \hbox{$\sim$}}} 1$
\cite{IngoldNato92in,KoenigCOTPRL97} 
start in the charge basis and treat tunneling as a weak perturbation,
while the semiclassical approach 
\cite{ZaikinSETJETP96,GeorgSETPRB98} starts from the
canonically conjugate phase representation and treats
small fluctuations  
about the classical paths which is adequate for
$\beta E_C 
 {\tiny{\lower 2pt \hbox{$<$}\atop\raise 5pt \hbox{$\sim$}}} 1$. 
So far no theoretical predictions are
available for the regime of highly conducting systems at low
temperatures which is within reach of state--of--the--art experiments.

In this work we study
the linear conductance of the SET for arbitrary parameters. 
Electron tunneling 
is treated nonperturbatively by means of a path integral 
Quantum Monte Carlo (QMC) approach employed to evaluate the 
current--current 
correlation function in imaginary time. 
We present an improved
singular value decomposition (SVD) scheme utilized to determine the 
conductance via an inverse Laplace transformation. This
allows us to obtain 
results for the entire range of parameters of experimental interest.

Specifically we consider
a SET consisting of two tunnel junctions with conductance $G_1$, $G_2$  
and capacitance $C_1$, $C_2$, respectively. The junctions in series are
biased by a voltage source $V$ and the
island between the tunneling barriers is coupled to a gate voltage $U_g$ 
via a capacitance $C_g$. From the Kubo formula, 
we write the dc conductance in terms of the 
current--current correlation function
$F(\tau)=\langle I(\tau) I(0) \rangle$ 
in imaginary time $\tau$ as \cite{GeorgSETPRB98}
\begin{equation}
 G
 =
 \lim_{\omega \rightarrow 0}
 \frac{1}{i \hbar \omega}
 \lim_{i \nu_n \rightarrow \omega  + i \delta}
  \int^{\hbar \beta}_0 \! d \tau\, e^{i \nu_n \tau} 
  F(\tau)  \,.
\label{eq:mcKubo}
\end{equation}
Once one has calculated the correlator $F(\tau)$ analytically, 
it is a rather easy task to continue its Fourier
transform to get the frequency dependent response. However, when
numerical techniques are used to evaluate $F(\tau)$, the
continuation becomes a serious problem: 
The Fourier transform is known only for a discrete set of equidistant
Matsubara frequencies $\nu_n=2\pi n/\hbar \beta$. Inserting the spectral 
representation of the correlation function
\cite{BaymGreensJMP61}
\begin{equation}
 F(\tau) 
 =
 \frac{1}{2\pi}
 \int_{-\infty}^\infty 
   d\omega \widetilde{F}(\omega) e^{-\tau \omega}
\label{eq:SpecDarst}
\end{equation}
with spectral density $\widetilde{F}(\omega) \ge 0$ into
$(\ref{eq:mcKubo})$, we find for the conductance
\begin{equation}
 G
 =
 \frac{\beta}{2} \widetilde{F}(0) \,.
\label{eq:mcconduct}
\end{equation}
To relate this with the numerically determined 
current--current correlation
function $F(\tau)$, one has to invert 
the two--sided
Laplace transformation $(\ref{eq:SpecDarst})$.
This operation, however, is known as an ill--posed problem:
small errors in $F(\tau)$ may cause large deviations in 
$\widetilde{F}(\omega)$. One has to use a 
regularization scheme to get rid of highly
oscillatory functions. Here, we employ 
SVD \cite{CreffieldSVDPRL95} regularized by
constraints explained below. Before addressing this, we present 
the QMC algorithm.

Explicitly, the current correlator may be written
\cite{GeorgSETPRB98}
\begin{equation}
 F(\tau)
 =
 4 \pi G_{\rm cl} \alpha(\tau)
 \langle \cos[\varphi(\tau)-\varphi(0)] \rangle
\label{eq:IICorrExpl}
\end{equation}    
where $G_{\rm cl}=G_1 G_2/(G_1+G_2)$ is the classical high temperature 
conductance. $\alpha(\tau)$ describes 
electron--hole pair propagation for electrons and holes 
created in different electrodes, and may be represented
in the form $(\ref{eq:SpecDarst})$
\begin{equation}
 \alpha(\tau) 
 =
 \frac{1}{2\pi}
 \int_{-\infty}^\infty d\omega 
  \widetilde{\alpha}(\omega) e^{-\tau \omega} 
\label{eq:mcalphaspec}
\end{equation}
with the spectral function 
\begin{equation}
 \widetilde{\alpha}(\omega) 
 =
 \frac{\hbar}{2\pi}
 \frac{\omega}{1-e^{-\hbar \beta \omega}} \,.
\label{eq:mcalphkernel}
\end{equation}
Finally, the average 
\begin{equation}
 A(\tau)
 =
 \langle \cos[\varphi(\tau)-\varphi(0)] \rangle 
\end{equation}
describes coupling between electron--hole pair excitations and the 
electromagnetic degrees of freedom, where the time derivative of the 
phase field $\varphi(\tau)$ is related to the voltage across the SET.
To calculate $A(\tau)$, we use a formally exact path integral
representation \cite{GeorgSETPRB98}
\begin{equation}
 A(\tau)
 =
 \frac{1}{Z}
 \int  D [\varphi] 
   e^{-\frac{1}{\hbar} S[\varphi]} 
   \cos[\varphi(\tau)-\varphi(0)]      \,,
\label{eq:pathgen}
\end{equation}
where $Z$ is the partition function and $S[\varphi]$ 
the Euclidian action that splits into
$S[\varphi]=S_C[\varphi]+S_T[\varphi]$. The first term
\begin{equation}
 S_C[\varphi]
 =
 \int_0^{\hbar\beta} d\tau
 \left[
  \frac{\hbar^2 \dot{\varphi}^2(\tau)}{4E_C}
  + i \hbar n_g \dot{\varphi}(\tau)
 \right]              \,,
\label{eq:coulac}
\end{equation}
where $E_C=e^2/2C$ is the charging energy with the island capacitance 
$C=C_1+C_2+C_g$, describes Coulomb charging in presence of a gate 
voltage $U_g= e n_g /C_g$. The second contribution
\begin{equation}
 \! S_T[\varphi]
 =
 2g
 \int_0^{\hbar\beta}  \!\! d\tau d\tau'
 \alpha(\tau-\tau') 
 \sin^2
 \left[
  \frac{\varphi(\tau)-\varphi(\tau')}{2}
 \right]
\label{eq:tunnelac}
\end{equation}
is due to electron tunneling where $g=(G_1+G_2)/G_K$ is
the dimensionless parallel conductance of the SET.

To implement the topology of the configuration space, we  
introduce winding numbers $k$ labeling
different classes of paths running from $\varphi(0)=0$ to 
$\varphi(\hbar \beta)= 2\pi k$. Changing the variables, 
$\varphi(\tau)= \nu_k \tau + \zeta(\tau)$ with 
$\zeta(0)=\zeta(\hbar \beta) = 0$, one may integrate out the gate voltage
dependent term in Eq.\ $(\ref{eq:coulac})$, and write 
\begin{equation}
 A(\tau)
 =
 \frac{
  \langle \langle 
     e^{-2\pi i k n_g} \cos[\nu_k \tau + \zeta(\tau)] 
  \rangle \rangle
      }{
  \langle \langle e^{-2\pi i k n_g} \rangle \rangle
       }      ,
\label{eq:averspec}
\end{equation}
with the unnormalized average over a positive measure
\begin{equation}
 \langle \langle X  \rangle \rangle
 =
 \sum_{k=-\infty}^\infty
 \oint  D [\zeta] 
   e^{-\frac{1}{\hbar} S_0[\nu_k \tau + \zeta]} X  \,.
\label{eq:patspec}
\end{equation}
Here, $S_0[\zeta]$ is given by Eqs.\ $(\ref{eq:coulac})$ and 
$(\ref{eq:tunnelac})$ with the gate voltage set to zero. 
Since $S_0[\zeta]$ is real, we may employ 
standard Metropolis sampling techniques
to evaluate the path integral numerically.
For the temperatures explored
the data were found to be saturated at Trotter number
$N\approx 5\beta E_C$. For higher 
temperatures we have chosen $N \ge 100$
to keep the resolution of the discretized correlation function 
high enough for the Laplace inversion.
The QMC data show statistical errors of at most 
$3\%$ for all temperatures and gate voltages. The denominator
$\langle \langle \exp(-2\pi i k n_g) \rangle \rangle$ in 
Eq.~$(\ref{eq:averspec})$ is an
averaged sign factor that always exceeds $0.04$ for the results shown 
below.

To analyse the data we start by noting that $A(\tau)$ 
also has a representation of the form
$(\ref{eq:SpecDarst})$. In terms of the spectral 
functions we then find
\begin{equation}
 \widetilde{F}(\omega)
 =
 2 G_{\rm cl} \int_{-\infty}^\infty d\omega'
 \widetilde{\alpha}(\omega -\omega') \widetilde{A}(\omega') \,.
\label{eq:mcconvolution}
\end{equation}
In view of the symmetry
$A(\tau) = A(\hbar \beta - \tau)$
equivalent to detailed balance
$
 \widetilde{A}(-\omega) 
 =
 e^{-\hbar \beta \omega} \widetilde{A}(\omega) \,,
$
we may introduce the positive and symmetric spectral
function 
\begin{equation}
 \widetilde{A}'(\omega)
 =
 \widetilde{A}(\omega) \, 
  \frac{1-e^{-\hbar \beta \omega}}{\omega}
\label{eq:mcsymmkernel}
\end{equation}
and write
\begin{equation}
 A(\tau)
 = 
 \frac{1}{2\pi}
 \int_0^\infty d\omega \widetilde{A}'(\omega)
   K_\beta(\omega,\tau)
\label{eq:mclaplacetrafo}
\end{equation}
with the integration kernel
\begin{equation}
 K_\beta(\omega,\tau)
 =
 \frac{\omega}{1-e^{-\hbar \beta \omega}} 
  \left[
   e^{-\omega \tau} + e^{-\omega (\hbar \beta-\tau)}
  \right]  .
\label{eq:mclaplacekernel}
\end{equation}
Now, we may restrict the time domain to 
$\tau \in [0,\hbar\beta/2]$. 
To invert the integral transform
$(\ref{eq:mclaplacetrafo})$ 
consider the operator ${\cal K}_\beta$ with kernel 
$K_\beta(\omega,\tau)$
\begin{equation}
 g(\tau)
 =
 ({\cal K}_\beta \widetilde{g})(\tau)
 =
 \int_{0}^{\omega_{\rm max}} d\omega 
   K_\beta(\omega,\tau) \widetilde{g}(\omega)  \,  ,
\label{eq:Operator}
\end{equation}
and it's adjoint ${\cal K}_\beta^*$
\begin{equation}
 \widetilde{f}(\omega)
 =
 ({\cal K}_\beta^* f)(\omega)
 =
 \int_0^{\hbar \beta/2} d\tau 
   K_\beta(\omega,\tau) f (\tau)  \,  .
\label{eq:AdjOperator}
\end{equation}
Clearly, the operation
${\cal K}_\beta{\cal K}_\beta^*$ is self--adjoint and positive. 
The eigenvalue equation
\begin{equation}
 ({\cal K}_\beta{\cal K}_\beta^* f_n)(\tau)= \mu_n^2 f_n(\tau)  
\end{equation}
can be solved numerically
by discretizing the corresponding integral equation. 
The resulting
eigenfunctions form an orthonormal basis $\{ f_n (\tau) \}$ 
in $L^2[0,\hbar \beta/2]$. Now,
the spectral function $\widetilde{A}'(\omega)$ can be expanded as
\begin{equation}
 \widetilde{A}'(\omega)
 \approx
 \sum_{n=1}^{n_{\rm max}} \frac{c_n}{\mu_n^2}  
 ({\cal K}_\beta^* f_n)(\omega) \, ,
\label{eq:SpecExpl}  
\end{equation}
where the 
\begin{equation}
 c_n 
 =
 \int_0^{\hbar \beta/2} d\tau A(\tau) f_n(\tau)
\label{eq:expcoeff}
\end{equation}
are expansion coefficients. Eq.~$(\ref{eq:SpecExpl})$ 
explicitly represents $\widetilde{A}(\omega)$ in terms of 
$A(\tau)$ and serves as 
inverse of the integral transform $(\ref{eq:mclaplacetrafo})$. 
Since with increasing $n$ the eigenvalues $\mu_n$ 
vanish overexponentially, 
for large $n$ highly oscillatory parts  
become more pronounced when the expansion
coefficients $c_n$ do not vanish fast enough. 
This exhibits the origin of the ill--posed problem: a small
error becomes amplified by a large inverse singular value 
$1/\mu_n^2$ and thus a regularization of the procedure is required.

First, the integral in Eq.~$(\ref{eq:Operator})$ is cut off at a
frequency $\omega_{\rm max}$  which is chosen so large that the
singular functions do not alter appreciably with this cutoff. 
Second, the sum in Eq.~$(\ref{eq:SpecExpl})$ is truncated 
at $n_{\rm max}$. 
To fix the value of $n_{\rm max}$ we use the noise level
\begin{equation}
 \sigma
 =
 \sqrt{\frac{1}{N+1} \sum_{i=0}^N \sigma(\tau_i)^2} 
\end{equation}
of the data
where $N$ is the Trotter number and 
$\sigma(\tau_i)$ the standard deviation of the data $A(\tau_i)$. 
This has to be compared with the 
error stemming from the truncated expansion in orthonormal functions 
\begin{equation}
 \epsilon_n
 = 
 \sqrt{\frac{1}{N+1}\sum_{i=0}^N
  \left( 
    \sum_{k=1}^{n}  c_k  f_k(\tau_i)-A(\tau_i)
  \right)^2}  \, .
\end{equation}
To ensure that we do not include fine structure originating from 
the noise of the QMC data, we fix $n_{\rm max}$ by 
demanding $\epsilon_{n_{\rm max}} \approx \sigma$.

An essential step
is to use additional information to regularize the SVD scheme.
We know that $\widetilde{A}'$ is positive, symmetric, vanishes
for large frequencies, and must obey the sum--rule
\begin{equation}
  \int_{-\infty}^\infty d\omega \widetilde{A}'(\omega)
    \frac{\omega}{1-e^{-\hbar \beta \omega}} 
  = 
  A(0) = 1 \,.
\end{equation}
To incorporate this, we make the ansatz
\begin{equation}
 \widetilde{A}_M'(\omega)
 = 
 \sum_{n=1}^M  
   \frac{a_n}{\mu_n^2} ({\cal K}^*_\beta f_n)(\omega)
\end{equation}
where $M>n_{\rm max}$. 
The coefficients $a_n$ are determined from minimizing a functional
according to three constraints. 
The {\it proximity} of the solution to the QMC--correlation function is
obtained by minimizing
\begin{equation}
 \Phi_{\rm prox}(a_1,...,a_{n_{\rm max}}) 
 = 
 \sum_{n=1}^{n_{\rm max}}
    \frac{(c_n - a_n)^2}{2\sigma_n^2} \,,
\label{eq:svdconstr2}
\end{equation}
where the $c_n$ are
the expansion coefficients $(\ref{eq:expcoeff})$ and where the points
are weighted according to errors in the QMC data.
To achieve {\it positivity} we minimize the integral
\begin{equation}  
 \Phi_{\rm pos}(a_1,...,a_M) 
 = 
 \int d\omega  \widetilde{A}_M'(\omega)^2 w(\omega)  \,,
\label{eq:svdconstr1}
\end{equation}
where the weight function vanishes in the region where the result
$(\ref{eq:SpecExpl})$ is positive.
To account for the {\it sum--rule} we minimize the functional
\begin{equation}  
 \Phi_{\rm sum}(a_1,...,a_M) 
 = 
 \left(1 - \int_{-\omega_{\rm max}}^{\omega_{\rm max}} d\omega  
   \frac{\omega \, \widetilde{A}_M'(\omega)}
        {1-e^{-\hbar \beta \omega}}\right)^2  .
\label{eq:svdconstr3}
\end{equation}
Overall we have to consider the linear combination
\begin{equation}
 \Phi(a_1,...,a_M) 
 = 
 \Phi_{\rm prox} + \lambda \Phi_{\rm pos} + \mu \Phi_{\rm sum} 
\label{eq:svdconstr}
\end{equation}               
with two regularization parameters $\lambda$ and $\mu$.   
Differentiating $(\ref{eq:svdconstr})$ with
respect to $a_n$ we get a system of linear equations
that may be solved. We have tested the method with 
analytical functions corrupted by Gaussian noise to optimize
the regularization parameters.

After extracting $\widetilde{A}'(\omega)$ from the QMC data via SVD,
we obtain $\widetilde{F}(\omega)$ from the convolution 
$(\ref{eq:mcconvolution})$ and the conductance from 
Eq.~$(\ref{eq:mcconduct})$.
Varying the gate voltage $U_g$ the conductance shows a periodic
behavior and we may define a maximum 
$G_{\rm max}=G|_{n_g=\scriptstyle{\frac{1}{2}}}$ and a minimum 
$G_{\rm min}=G|_{n_g=0}$ conductance in dependence on temperature. 

To estimate error bars we add a further
functional to $(\ref{eq:svdconstr})$: We try to minimize the 
distance of $\widetilde{F}(\omega)$
to a fictive either too low or too high value 
$\widetilde{F}^\pm(\omega)$. 
This fictive value of the conductance
still has to be in accordance with the other three constraints
and leads to an upper and lower bound of possible conductance
\cite{HuepperSVD99}.
The error bars for the conductance found by this procedure lie all 
below $3\%$. Since they are of the order of the symbol sizes they
will be omitted in the figures.

\begin{figure}[t]
\begin{center}
\leavevmode
\epsfxsize=0.45 \textwidth
\epsfbox{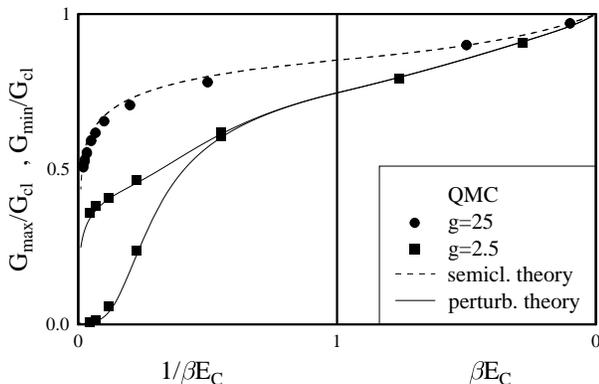}
\vspace*{-.0cm}
\caption{Predictions for the maximum and minimum linear conductance 
of a SET for dimensionless parallel 
conductance $g=2.5$ compared to second order perturbative
results \protect\cite{KoenigCOTPRL97} and for $g=25$ compared to 
semiclassical findings \protect\cite{ZaikinSETJETP96,GeorgSETPRB98}.
}
\label{fig:setnth}
\end{center}
\end{figure}

As a first test of the approach we compare our numerical data 
with second order perturbative results 
\cite{KoenigCOTPRL97} and with
semiclassical findings 
\cite{ZaikinSETJETP96,GeorgSETPRB98}.
The comparison in Fig.~\ref{fig:setnth} shows that our results 
agree well with both limiting analytical theories in their ranges of
applicability and therefore 
bridge between the regions of parameters covered by analytical work.

Recently, P.\ Joyez {\it et al}.\ \cite{JoyezSETPRL97} and
D.\ Chouvaev {\it et al}.\ \cite{KuzminSETPRB99} have measured 
the conductance of SETs in the moderately--to--strong tunneling 
regime. The strong tunneling data by Chouvaev {\it et al}.\ 
cover only a restricted range of temperatures and were found  
to be in agreement with semiclassical theory. 
In contrast, some of the data obtained by Joyez {\it et al}.\
are not only outside of the range of the perturbative theory but also 
explore temperatures low enough to show strong deviations form
semiclassical results.

\begin{figure}[t]
\begin{center}
\leavevmode
\epsfxsize=0.45 \textwidth
\epsfbox{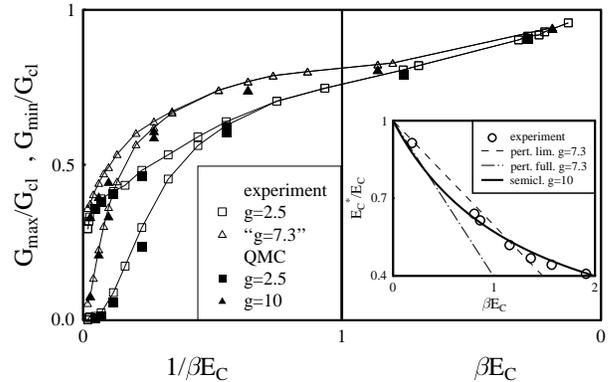}
\vspace*{-.0cm}
\caption{Predictions for the maximum and minimum linear conductance 
of a SET for dimensionless parallel 
conductance $g=2.5$ and $g=10$ compared with experimental findings 
from Ref.~\protect\cite{JoyezSETPRL97}. In the inset the experimental
data of the renormalized charging energy $E_C^*/E_C$
for ``$g=7.3$'' are shown and
compared with high temperature approximations (see main text).
}
\label{fig:setn}
\end{center}
\end{figure}

However, the experimental data in Ref.~\cite{JoyezSETPRL97} 
for strong tunneling ``$g=7.3$'' are found to deviate significantly 
from our predictions for $g=7.3$ (data not shown). 
Therefore, we have reexamined the parameters given in
Ref.~\cite{JoyezSETPRL97}: The bare 
charging energy $E_C$ was determined independently 
from resonances in the superconducting 
state that should be reliable within errors of about $10\%$. 
The series conductance was measured at 
sufficiently high temperatures ($\beta E_C = 0.18$) fixing 
$G_{\rm cl}$ up to a small error.
Assuming identical tunnel junctions, the dimensionless 
parallel conductance was estimated as
$g=4 G_{\rm cl}/G_K$.
This assumption was consistent with a fit
(dashed line in inset of Fig.~\ref{fig:setn})
of the high temperature data
to the prediction 
$
 G/G_{\rm cl}
 =
 1-\beta E_C^*/3 
$ 
with 
\begin{equation}
 E_C^*/E_C
 = 
 1- (9 g \zeta(3) / 2\pi^4 + c) \beta E_C 
\label{eq:mcecren}
\end{equation}
where $c=0$. Yet, the correct result reads $c=1/5$ 
\cite{GeorgSETPRB98,KoenigZaikinPRB97} leading to the 
dashed--dotted line in the inset which is inconsistent 
with the data. Even for the highest temperatures explored
experimentally, the data should be compared with the full 
semiclassical prediction from Ref.~\cite{GeorgSETPRB98}. 
As can be seen from the inset in Fig.~\ref{fig:setn},
the high temperature data are only consistent with a parallel
conductance near $g=10$ implying an asymmetry 
of the transistor.

\vspace*{-0.3cm}
\vspace*{-0.3cm}

In Fig.~\ref{fig:setn} we show for $g=10$ QMC predictions
for maximum and minimum linear 
conductance in dependence on the inverse temperature. 
They are compared with the
experimental findings from Ref.~\cite{JoyezSETPRL97}
for ``$g=7.3$'' and are found to be in good agreement.
The error bounds of the numerical data are again within 
the symbol size. The parameters for the QMC were not adjusted to  
improve the global fit but coincide with those extracted from the high
temperature analysis. Remaining deviations between the experimental
data and our predictions may
result from the uncertainty of the parameters.

In summary we have presented a general approach to calculate the 
conductance of the SET for arbitrary parameters. We found good
agreement with limiting analytical theories and could explain 
experimental findings for highly transmitting
tunnel junctions. As one of the results of this study future
experiments in the strong tunneling regime should explore the high
temperature behavior of the SET more carefully to allow for a unique 
parameter identification within the range of validity of the
semiclassical theory.

\vspace*{-0.3cm}
\vspace*{-0.4cm}

The authors would like to thank the authors of 
Ref.~\cite{JoyezSETPRL97} for valuable discussions.
One of us (GG) acknowledges hospitality of the CEA-Saclay 
during an extended stay.
Financial support was provided by the Deutsche
Forschungsgemeinschaft (DFG), the Deutscher Akademischer 
Austauschdienst (DAAD), and the MINERVA Foundation (Munich).

\end{multicols}
\end{document}